\documentclass{aa}
\usepackage{txfonts}
\usepackage{psfig}
\usepackage{wasysym}
\begin{document}

\newcommand{\beqa}{\begin{eqnarray*}}
\newcommand{\eeqa}{\end{eqnarray*}}
\newcommand{\beqan}{\begin{eqnarray}}
\newcommand{\eeqan}{\end{eqnarray}}
\newcommand{\beq}{\begin{equation}}
\newcommand{\eeq}{\end{equation}}
\newcommand{\diff}{{\rm d}}
\newcommand{\drr}{\frac{\partial}{\partial r}}
\newcommand{\dtt}{\frac{\diff}{\diff t}}
\newcommand{\dr}[1]{\frac{\partial  #1}{\partial r}}
\newcommand{\dt}[1]{\frac{\partial  #1}{\partial t}}
\newcommand{\lp}{ \left(}
\newcommand{\rp}{ \right)}
\newcommand{\lc}{ \left[}
\newcommand{\rc}{ \right]}
\newcommand{\cf}{{\it cf.}~}
\newcommand{\ie}{{\it i.e.}~}
\newcommand{\eg}{{\it e.g.}~}
\newcommand{\moyh}[1]{\left<  #1  \right>_{h}}
\newcommand{\moyl}[1]{\left<  #1  \right>_{\theta}}
\newcommand{\moyhc}[1]{\left< \left|  #1  \right| ^2 \right> _{h}}
\newcommand{\moylc}[1]{\left< \left|  #1  \right| ^2 \right> _{\theta}}

\def\O{\Omega}

\bibliographystyle{plain}

\title{Angular momentum transport by internal gravity waves}
\subtitle{III - Wave excitation by core convection and the Coriolis effect}

\author{Florian P. Pantillon\inst{1,2}, Suzanne Talon\inst{2}, \and Corinne Charbonnel\inst{1,3}}

\offprints{Suzanne Talon}

\institute{
Geneva Observatory, University of Geneva, ch. des Maillettes 51, 1290 Sauverny, Switzerland 
\and D\'epartement de Physique, Universit\'e de Montr\'eal, Montr\'eal PQ H3C 3J7, Canada
\and Laboratoire d'Astrophysique de Toulouse et Tarbes, CNRS UMR 5572, OMP, Universit\'e Paul 
Sabatier 3, 14 Av. E.Belin, 31400 Toulouse, France \\
(pantillon@astro.umontreal.ca, talon@astro.umontreal.ca, Corinne.Charbonnel@obs.unige.ch)}

\date{Received / Accepted }

\authorrunning{Pantillon, Talon \& Charbonnel}
\titlerunning{Internal gravity waves and the Coriolis effect}

  \abstract
   {This is the third in a series of papers that deal with angular momentum 
    transport by internal gravity waves.
    We concentrate on the waves excited by core convection in a $3\,M_{\odot}$, Pop I 
    main sequence star.}
   {Here, we want to examine the role of the Coriolis acceleration in the equations
   of motion that describe the behavior of waves and to evaluate its impact
   on angular momentum transport.}
   {We use the so-called traditional approximation of geophysics, which allows 
   variable separation in radial and horizontal components. In the presence of
   rotation, the horizontal structure is described by Hough functions instead of
   spherical harmonics. }
   {The Coriolis acceleration has two main effects on waves. It transforms pure gravity
   waves into gravito-inertial waves that have a larger amplitude closer to
   the equator, and it introduces new waves whose restoring force is mainly the 
   conservation of vorticity.}
   {Taking the Coriolis acceleration into account changes the subtle balance 
   between prograde and retrograde waves in non-rotating stars. It
   also introduces new types of waves that are either purely prograde or
   retrograde. We show in this paper where the local deposition of angular 
   momentum by such waves is important.}

   \keywords{Hydrodynamics -- Turbulence -- Waves --
              Methods: numerical -- Stars: interiors -- Stars: rotation
               }

   \maketitle

\section{Introduction}

Internal gravity waves (IGWs) have received growing attention in the past 15 years
as a source of angular momentum redistribution, after the suggestion by various
authors that they could be responsible for the Sun's quasi-flat rotation
profile (Schatzman~1993; Zahn, Talon \& Matias~1997; Kumar \& Quataert~1997).
Charbonnel \& Talon~(2005) convincingly show
that IGWs can play a major role in
carrying angular momentum from the rapidly rotating core, left after the Sun's
original contraction, to the surface convection zone that is 
continuously spun down by magnetic braking. In self-consistent
evolutionary models that also takes into account the transport of angular momentum
and chemical species by meridional circulation and shear turbulence, 
they also showed that the
associated destruction of lithium is compatible with the destruction measured in the
Sun. 
In the case of F dwarfs, Talon \& Charbonnel~(2003) show that the combination
of surface breaking for stars with an effective temperature $T_{\rm eff}
\apprle 6900\,{\rm K}$ and the appearance of IGWs at a temperature
$T_{\rm eff} \apprle 6600\,{\rm K}$ when the surface convection zone is thick
enough can together explain the existence of the so-called lithium dip discovered by
Wallerstein et al. (1965).

Now the question arises as to whether IGWs 
generated by a convection core 
also play a role in the evolution of massive stars\footnote{By massive 
stars we mean objects for which central hydrogen burning occurs in a convection core,
{\it i.e.}, stars with $M_{\rm ini} \apprge 2.2\,M_{\odot}$ at solar metallicity.}.
Indeed, rotational mixing is now admittedly a major ingredient in
such objects (see \eg Maeder \& Meynet~2000; Heger, Woosley \& 
Langer~2000), and the existence of another major source of angular momentum
redistribution within the star must be examined.

However, as a general rule, massive stars are fast rotators. Here, fast must
be understood in comparison with the wave frequencies that dominate angular
momentum redistribution in stars; that is, $\sigma \approx 1\,\mu{\rm Hz}$ (Talon
\& Charbonnel~2005). Our goal in this paper is to establish how wave properties
are altered by rotation in preparation for full calculations of angular momentum
redistribution in massive stars. \S\,\ref{sec:eigenf} is devoted to 
examining the (horizontal) eigenfunctions in the presence of rotation,
while \S\,\ref{sec:angmom} presents the results on angular momentum transport
itself.

\section{Eigenfunctions in rotating stars \label{sec:eigenf}}

\subsection{The traditional approximation}
In the study of internal waves in the presence of rotation,
two effects should be considered: the star's distortion caused by the
centrifugal force, and the modification to the momentum equation by the Coriolis
acceleration. In 1--D modeling, the stellar distortion can be accounted for by
calculating an average centrifugal force that locally reduces gravity. This effect
is quite important at the stellar surface where the ratio $r^3 \Omega^2/GM_r$ is
at its maximum, and it modifies the location of the star in the HR diagram
(see \eg the ZAMS position of stars of various velocities in Fig.~5 of 
Talon et al.~1997).
It does however have a minor
impact on stellar evolution, since it modifies the physical
conditions near the stellar core only slightly 
(see \eg Maeder \& Meynet 2000). 
In this work, we will neglect this term.
The Coriolis acceleration, on the other hand, has a direct effect on the 
oscillations and modifies
the momentum equations for the displacement $\vec{\xi}$, which become
\beq
-\sigma ^2 \xi_r - \underbrace{2i\sigma \Omega \sin \theta \, \xi_\phi}=
-\frac{1}{\rho}\frac{\partial}{\partial r} P' -  \frac{\rho'}{\rho}g \label{momentum1} 
\eeq
\beq
-\sigma ^2 \xi_\theta - 2i\sigma \Omega \cos \theta \, \xi_\phi=
-\frac{1}{\rho r}\frac{\partial}{\partial \theta} P' \label{momentum2} 
\eeq
\beqan
\lefteqn{
-\sigma ^2 \xi_\phi 
+\underbrace{ i\sigma \lp r \frac{\diff \Omega}{\diff r}  
+ 2 \Omega \rp \sin \theta \, \xi_r } 
+ 2i\sigma \Omega \cos \theta \, \xi_\theta} \label{momentum3} \\
&& \hspace{5cm} = -\frac{1}{\rho r \sin \theta }\frac{\partial}{\partial \phi}
P'\nonumber 
\eeqan
where $\sigma$ is the wave frequency, and other symbols have their usual meaning. 
Here, we have neglected variations in the gravitational potential ($\box0{\Phi'=0}$, 
Cowling 1941), which is particularly justified for gravity waves 
that have mostly horizontal displacements.
This equation was given in Talon~(1997), and it differs from the one used by
other authors (see below) only by the derivative in $\Omega$.

\begin{figure}
\centerline{
\psfig{figure=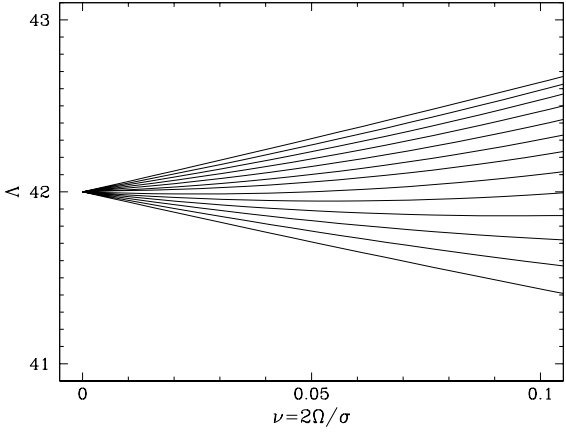,height=6cm,angle=-90}
}
\caption{Eigenvalues of Laplace's tidal equation when rotation is slowly 
increased. Illustrated
modes correspond to the $\ell =6$ non-rotating case.
\label{fig:degeneracy}}
\end{figure}

To proceed further, we make a simplification in our system of 
equations~(\ref{momentum1}, \ref{momentum2}, \ref{momentum3}) 
and neglect the horizontal component of the
rotation vector $\Omega_h=\Omega \sin \theta$. In the case of Eq.~($\ref{momentum1}$), this is
justified by comparing this term with the vertical gradient of the pressure perturbation. 
In Eq.~($\ref{momentum3}$), we
invoke the dispersion relation of gravito-inertial waves (see \eg Unno et al. 1989)
\beq
\sigma ^2 = \frac{N^2 k_h^2 + \lp 2 \vec{\Omega} \cdot \vec{k} \rp ^2 }{k^2},
\eeq
which in the case $\sigma^2, (2\Omega)^2 \ll N^2$ (the condition for the
approximation to be valid), yields
\beq
k_r^2 \gg k_h^2. \label{wavenumber}
\eeq
This is similar to the result obtained for pure gravity waves. Since these low-frequency 
oscillations are quasi-incompressible, Eq.~(\ref{wavenumber}) is equivalent to
\beq
\xi_h^2 \gg \xi_r^2,
\eeq
whence the neglect of the second term ($\propto \xi_r$) of Eq.~(\ref{momentum3}).
This is known as the {\em traditional approximation}, which is well known in 
geophysics (see \eg Eckart~1960), and it was used for the first time in an
astrophysical context by Berthomieu et al.~(1978). 

\begin{figure}
\centerline{
\psfig{figure=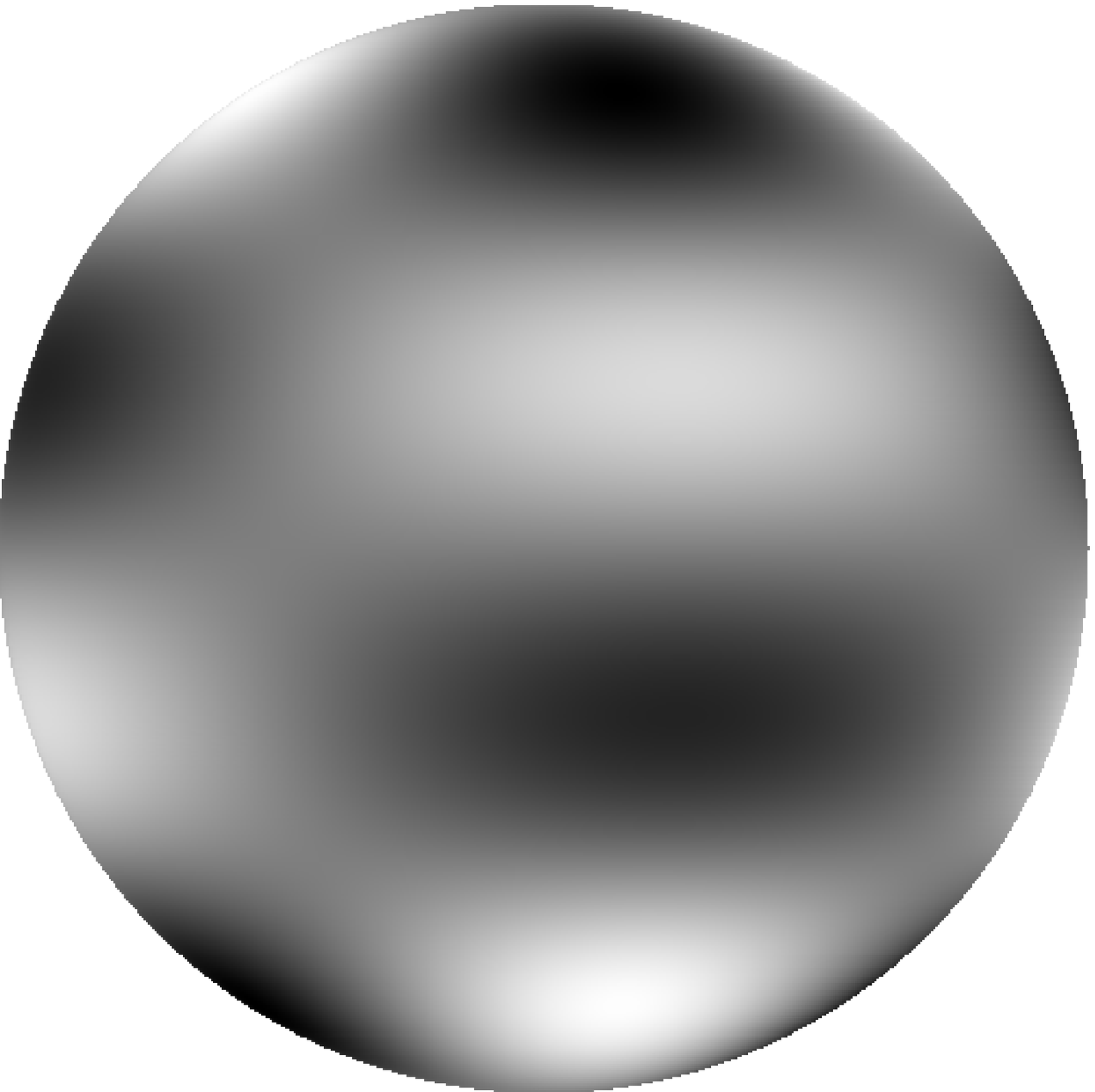,height=6cm}
}
\vspace*{0.2cm}
\centerline{
\psfig{figure=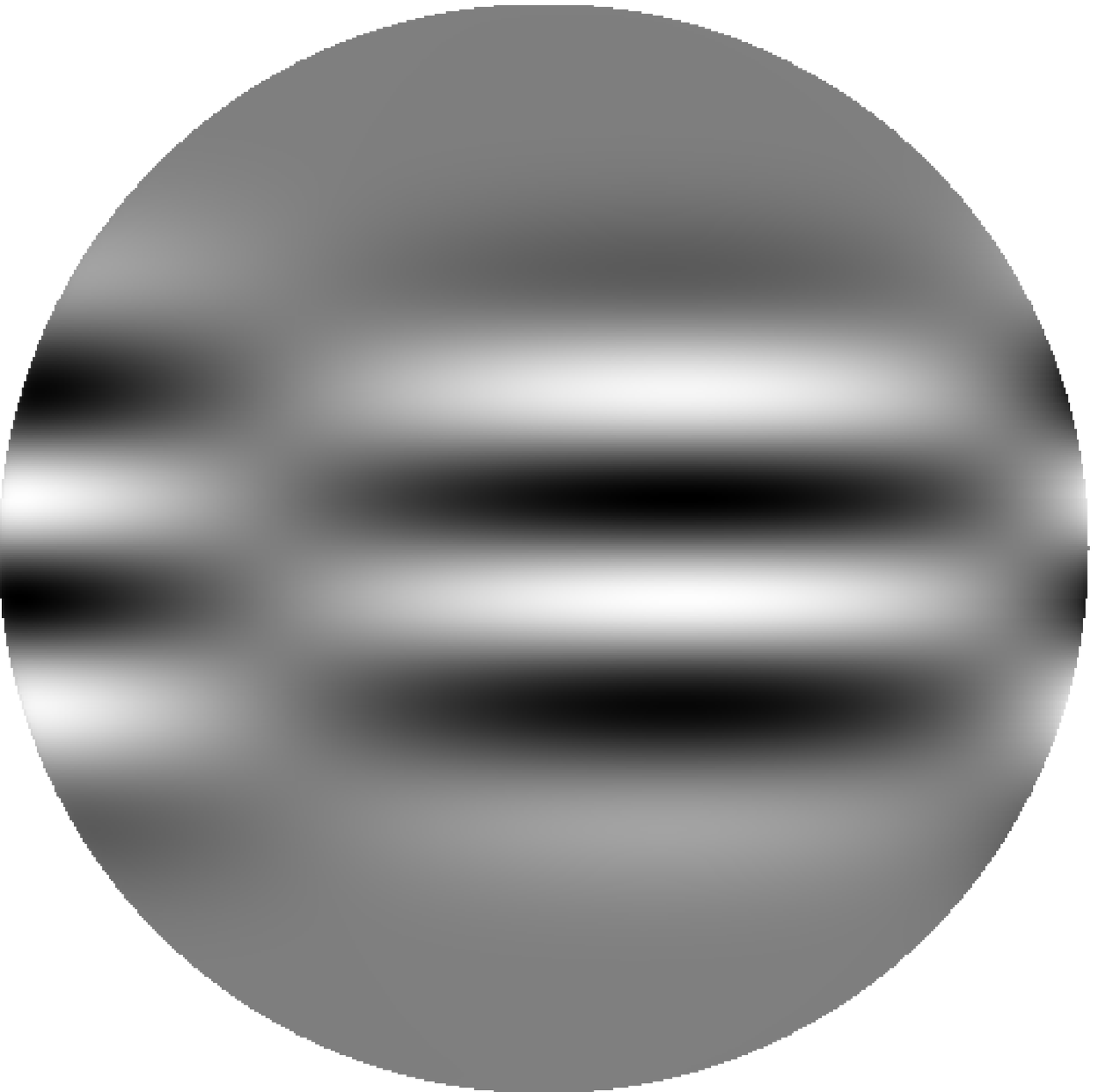,height=6cm}
}
\caption{{\bf Top:} Spherical harmonic $Y_\ell ^m$ with $\ell=5$, $m=+2$
{\bf Bottom:} Corresponding Hough function $\Theta_{km} e^{im\phi}$ with $k=+3$, $m=+2$ and
$\nu \approx 2$.
\label{fig:modes}}
\end{figure}

In this case, wave eigenfunctions can be separated into radial and horizontal components.
For the radial displacement, for example, one gets\footnote{Note that here 
we are using the convention of negative $m$ for prograde modes, contrary to 
what has been used in previous papers.}
\beq
\xi_r \lp r,\theta,\phi,t \rp = \xi_r \lp r\rp \Theta\lp \theta \rp e^{im\phi}e^{i\sigma t}.
\label{xiradial}
\eeq
The longitudinal eigenfunction is no longer given by an associate Legendre polynomial, as in the 
non rotating limit, but obeys instead a new differential eigenvalue problem of
the form
\beq
{\cal L}_\nu \lc \Theta \lp \theta \rp \rc = - \Lambda \Theta \lp \theta \rp ,
\eeq
where
\beqan
\lefteqn{{\cal L}_\nu = \frac{\diff}{\diff x} \lp \frac{1-x^2}{1-\nu^2x^2} \frac{\diff}{\diff x} \rp} 
\label{laplace}\\
&& \hspace*{2cm}- \frac{1}{1-\nu^2x^2} \lp \frac{m^2}{1-x^2} + m\nu \frac{1+\nu^2x^2}{1-\nu^2x^2}\rp.
\nonumber 
\eeqan
In this equation, we use the rotation parameter 
\beq
\nu \equiv \frac{2\Omega}{\sigma}
\eeq
and define $x=\cos \theta$. The azimuthal order takes
integer values $m=0, \pm 1, \pm 2, \dots$
This equation, known as {\em Laplace's tidal equation}, 
was originally derived in the context of terrestrial tides. It forms a
Sturm-Liouville problem, ensuring the existence of a base of
orthogonal eigenfunctions.
These were described in the geophysical context by
Hough~(1898), and the first numerical solution was calculated by Longuet-Higgins~(1968).
More recently, they have been examined in the context of neutron stars (Bildsten,
Ushomirsky \& Cutler~1996) and in studies of low-frequency oscillations 
in rotating stars (Lee \& Saio~1997; Daszy\'nska-Daszkiewicz,
Dziembowski, \& Pamyatnykh~2007 and references therein). 
We wrote a relaxation code to 
solve this equation numerically (see Press et al.~1992). A similar procedure is
also described in Lee \& Saio~(1997).

The Hough function $\Theta_{km} \lp \nu, \cos \theta \rp$ 
is associated with the eigenvalue
$\Lambda _{km}\lp \nu \rp$. Following Lee \& Saio~(1997), 
the order $k=0, \pm 1, \pm 2, \dots$, which does not explicitly appear in
Eq.~(\ref{laplace}), is such that $\Lambda_{k+1,m}\lp \nu \rp > \Lambda_{km}\lp \nu \rp$.
In the $\nu=0$ limit, Eq.~(8) is identical to Legendre's equation
with $k=\ell - \left| m \right|$ and $\Lambda_{km}=\ell \lp \ell + 1 \rp $.
Slowly increasing $\nu$, we first lift the degeneracy in $\Lambda$ 
(see Fig.~\ref{fig:degeneracy}). Increasing rotation confines eigenmodes toward
the stellar equator (see Fig.~\ref{fig:modes}).

In the case of more rapid rotation ({\it i.e.}, $\sigma \apprle 2\Omega$),
mode organization becomes much less trivial.
First, new solutions appear when $\nu>1$ and have negative eigenvalues; thus they
are only evanescent. However, as $\nu$ increases, so do their eigenvalues, which become
positive for the retrograde modes (and so the waves become propagative) when 
$\nu = \lp m-k \rp \lp m-k-1\rp/m$ (\cf Lee \& Saio~1997).
Second, a large fraction of the modes that exist in the absence of rotation see their
eigenvalue grow tremendously with rotation (see Fig.~\ref{fig:lambda}).
Townsend~(2003) has introduced a new classification for modes based on the behavior
of their eigenvalues in rapid rotation:
\begin{figure}
\centerline{
\psfig{figure=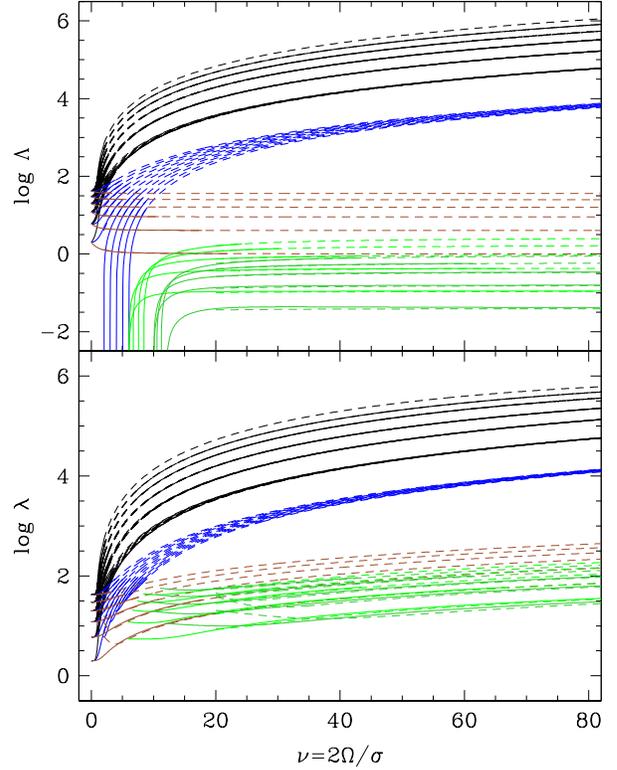,height=11cm}
}
\caption{Eigenvalues $\Lambda$ of Laplace's tidal equation 
for rapid rotation (top) and equivalent eigenvalues $\lambda$ 
for the horizontal Laplacian of Hough functions (bottom).
Mode classification: {\bf Black:} Gravito-inertial waves {\bf Blue:} Yanai waves
{\bf Brown:} Kelvin waves {\bf Green:} Rossby waves. Continuous lines correspond
to numerical solutions and dashed lines
to asymptotic solutions (see \S\,\ref{sec:asym}). 
\label{fig:lambda}}
\end{figure}
\begin{figure*}
\centerline{
\psfig{figure=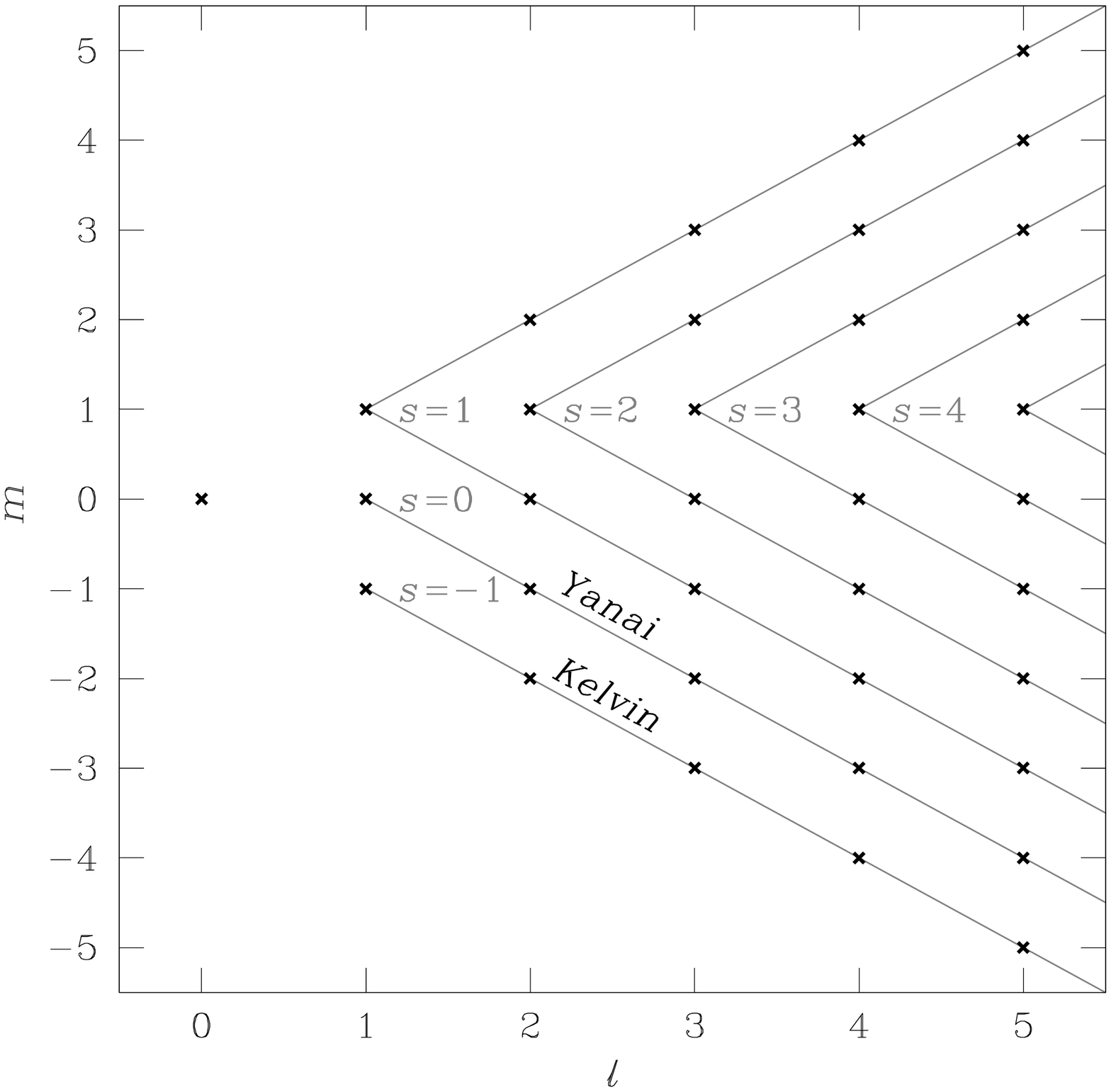,height=7.5cm}
\psfig{figure=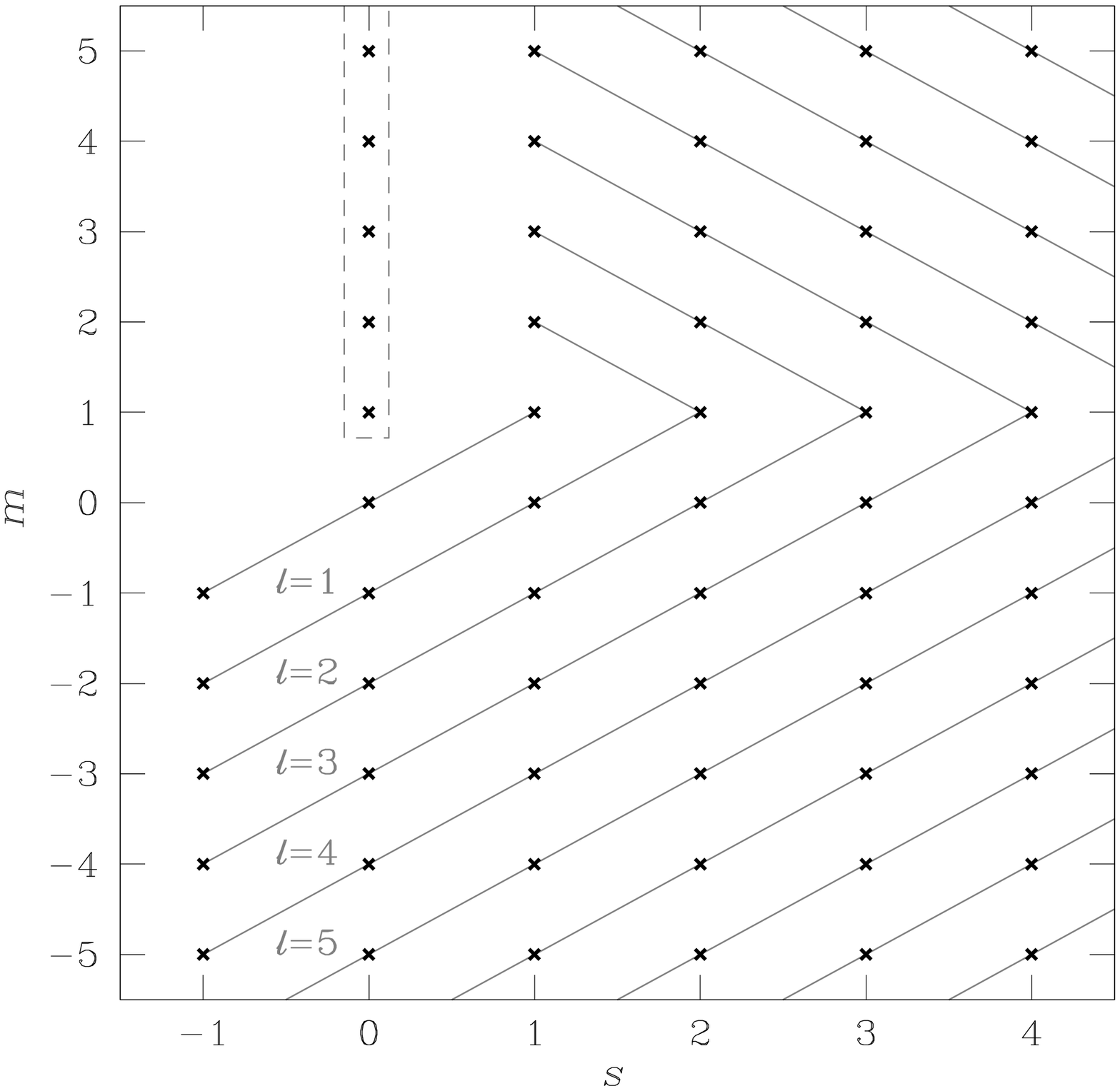,height=7.5cm}
}
\caption{Correspondence of index $s$ with index
$\ell$ in the absence of rotation. Modes in the dashed box (retrograde Yanai waves)
do not exist in the absence of rotation.}
\label{fig:classification}
\end{figure*}
\begin{itemize}
\item {\em Gravito-inertial waves}: they are similar to normal gravity waves, but
modified by the Coriolis acceleration. They are assigned indexes $s=1,2,3,\dots$,
such that $\Lambda_{s+1,m}\lp \nu \rp > \Lambda_{sm}\lp \nu \rp$.
\item {\em Rossby waves}: these are purely retrograde waves, which exist only in the case of rapid rotation. 
They arise from the conservation of specific vorticity,
combined with the effect of curvature. 
They also have indexes $s=1,2,3,\dots$ but ordered such that 
$\Lambda_{s+1,m}\lp \nu \rp < \Lambda_{sm}\lp \nu \rp$.
The $s=1$ modes are retrograde sectoral waves.
\item {\em Yanai waves\footnote{They have been named after their discoverer in the Earth's
atmosphere, Yanai \& Maruyama~(1966).}}: they behave 
like a mixture of gravity and Rossby waves. The $m \le 0$ modes exist in 
the absence of rotation. The $m>0$ modes appear when $\nu=m+1$ with small eigenvalues, and their
horizontal eigenfunctions are then exactly $\Theta\lp \nu = m+1 \rp = P_{m+1}^m$.
When they appear and have small eigenvalues, they behave mostly like
Rossby waves; $m \le 0$ waves and $m>0$ waves with large eigenvalues
behave rather like gravity waves.
We assign to them index $s=0$. 
\item {\em Kelvin waves}: like Rossby waves, they arise from
the conservation of specific vorticity combined this time with 
the stratification of the medium.
They are purely prograde waves, whose characteristics change little with rotation.
This occurs because their displacement is very small in the $\theta$ direction.
They correspond to the $m<0$ sectoral modes.
We assign to them index $s=-1$.
\end{itemize}
Without rotation, the correspondence between various indexes follows
\beq
s=\ell-m+1 ~~{\rm for}~m>0 
\eeq
and
\beq
s=\ell+m-1 ~~{\rm for}~m\le0
\eeq
and is illustrated in Fig.~\ref{fig:classification}.

\begin{figure*}
\centerline{
\psfig{figure=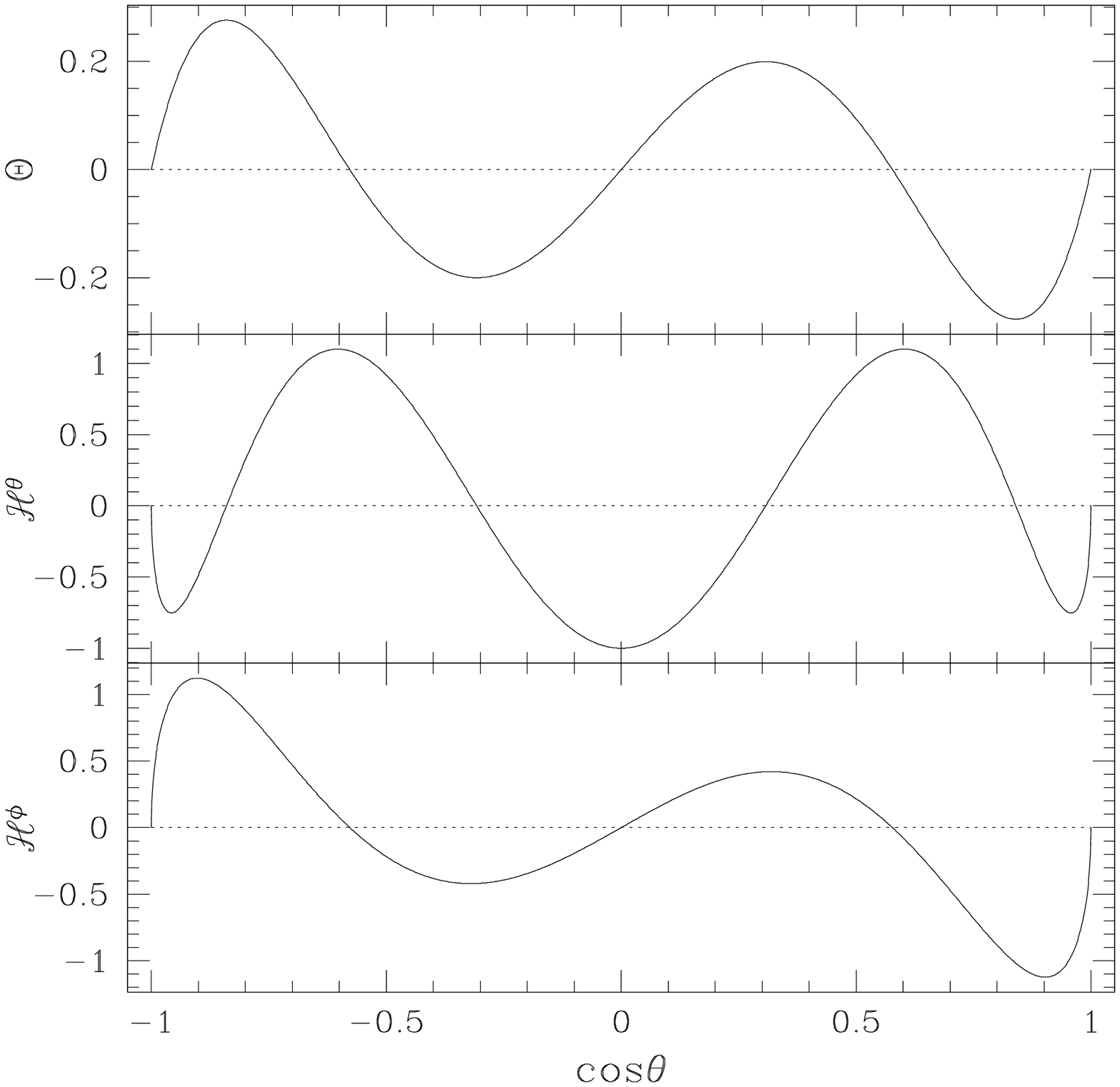,height=7.5cm}
\psfig{figure=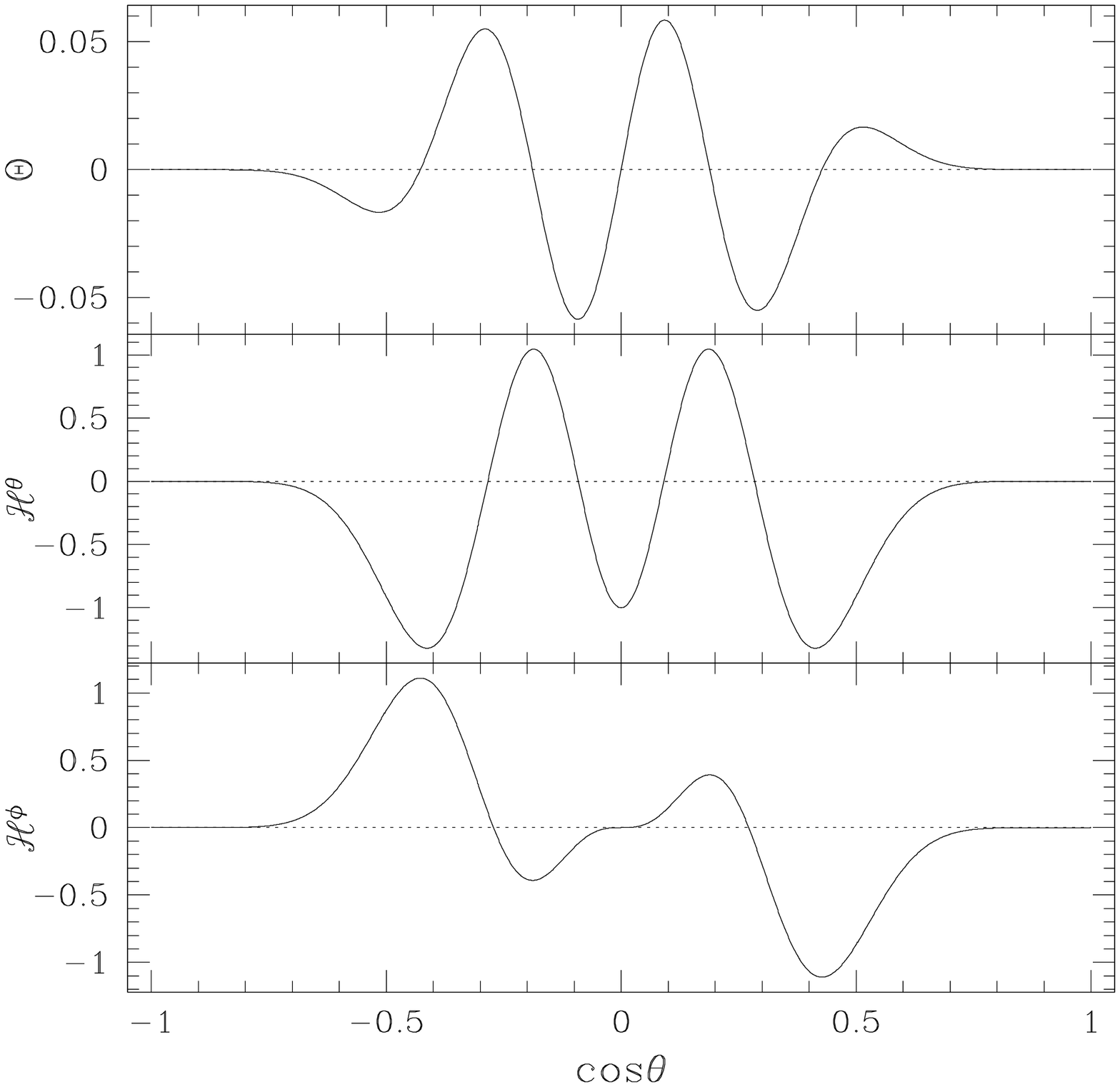,height=7.5cm}
}
\caption{
Hough functions $\Theta_{sm}$, ${\cal H}^\theta_{sm}$, and
${\cal H}^\phi_{sm}$. Normalization is given by 
$\diff \Theta /\diff \cos \theta=1$ at $\cos \theta =0$.
{\bf Left:} $s=4$, $m=2$, and $\nu=0$ ($\ell = 5$). In this case, 
$\Theta_{sm}$, ${\cal H}^\theta_{sm}$, and ${\cal H}^\phi_{sm}$ are given by 
$P_\ell ^m$, $\diff P_\ell ^m/\diff \theta$, and $mP_\ell ^m /\sin \theta$ respectively.
{\bf Right:} $s=4$, $m=2$, and $\nu \approx 2$. 
\label{fig:hough}}
\end{figure*}

\subsection{Horizontal displacement}
Pressure ($P'$) and density ($\rho'$) perturbations have a horizontal
structure identical to that of the radial displacement (Eq.~\ref{xiradial}).
This is, however, not the case for the two components of the horizontal
displacement, which take the form
\beq
\xi_\theta \lp r,\theta,\phi,t \rp = \frac{1}{r\sigma^2} \frac{P'(r)}{\rho}
{\cal H} ^\theta \lp \nu ,\cos \theta \rp e^ {im\phi} e^{i\sigma t}
\eeq
\beq
\xi_\phi \lp r,\theta,\phi,t \rp = \frac{i}{r\sigma^2} \frac{P'(r)}{\rho}
{\cal H} ^\phi \lp \nu ,\cos \theta \rp e^ {im\phi} e^{i\sigma t}
\eeq
with
\beqan
\lefteqn{{\cal H} ^\theta _{sm} \lp \nu ,\cos \theta \rp =
\frac{1}{\lp 1-\nu^2x^2\rp \sqrt{1-x^2}}} \nonumber \\
&&\hspace*{1.5cm} \times \lc -\lp 1-x^2 \rp \frac{\diff}{\diff
x} + m\nu x \rc \Theta_{sm} \lp \nu ,\cos \theta \rp
\eeqan
\beqan
\lefteqn{{\cal H} ^\phi _{sm} \lp \nu ,\cos \theta \rp =
\frac{1}{\lp 1-\nu^2x^2\rp \sqrt{1-x^2}}} \nonumber \\
&&\hspace*{1.5cm} \times \lc -\nu x\lp 1-x^2 \rp \frac{\diff}{\diff
x} + m \rc \Theta_{sm} \lp \nu ,\cos \theta \rp
\eeqan
where again $x=\cos \theta$. An example of these functions is shown in 
Fig.~\ref{fig:hough}. Note that $\Theta_{sm}$ 
has two more zeros compared to the non-rotating case. Furthermore, 
${\cal H}^\theta_{sm}$ and ${\cal H}^\phi_{sm}$ are no longer given 
by simple expressions and have zeros that do not coincide
with those of $\Theta_{sm}$.

\subsection{Asymptotic solution \label{sec:asym}}
When considering the impact of IGWs on the distribution of angular momentum 
within a star, low-frequency waves (with $\sigma \approx 1 \,\mu {\rm
Hz}$) play a dominant role (this is discussed 
at length in Talon \& Charbonnel~2005). In the case of massive stars
that are generally rapid rotators (with $\Omega \approx 20\,\mu {\rm
Hz}$), we are within the limit $\nu \gg 1$, where
asymptotic solutions to Laplace's equation exist
(Townsend~2003).

To get these solutions, Townsend defines
\beq
\widehat{\Theta} \lp \cos \theta \rp = \sin \theta \, {\cal H}^\theta \lp \cos \theta \rp
\eeq
\beq
\widetilde{\Theta} \lp \cos \theta \rp = -\sin \theta \,{\cal H}^\phi \lp \cos \theta \rp.
\eeq
Then, eliminating $\widetilde{\Theta}$, Laplace's equation is equivalent
to two coupled first-order equations
\beq
\lc \lp 1-x^2 \rp \frac{\diff}{\diff x} - m \nu x \rc \Theta
=\lp \nu^2 x^2 -1 \rp \widehat{\Theta}
\eeq
\beq
\lc \lp 1-x^2 \rp \frac{\diff}{\diff x} + m \nu x \rc \widehat{\Theta}
=\lc\Lambda \lp 1-x^2 \rp -m^2 \rc \Theta.
\eeq

Further simplifications to these equations can be made.
For {\em gravito-inertial waves} and {\em Yanai waves}, $\Lambda \gg m^2$,
while {\em Rossby waves} have $\Lambda \ll m^2$ (see Fig.~\ref{fig:lambda}). 
When $\nu \gg 1$, these waves are very confined to the equator, which
allows the assumption $\lp 1-x^2\rp \approx 1$. For {\em Kelvin waves},
one has $\Lambda \approx m^2$, leading to $| \Theta | \gg | \widehat{\Theta} |$.
Townsend~(2003) then finds approximate solutions to the Hough functions
in the form of Hermite polynomials. We refer the reader to his paper for
their derivation. Let us note here that, except for Rossby waves that are less
confined to the equator, a direct comparison with the numerical solution for the
complete version of Laplace's equation shows that the approximation is valid
for $\nu \apprge 2$. For Rossby waves, this approximation becomes valid
at much higher values on the order of $\nu \apprge 20$
(see Fig.~\ref{fig:lambda}).

In this asymptotic limit, eigenvalues take the form
\beq
\Lambda_{sm}\lp \nu \rp = \lc \frac{1}{2} \nu \lp 2s+1 \rp + \frac{1}{2}
\sqrt{\nu^2 \lp 2s+1 \rp ^2 -4 \lp m\nu - m^2 \rp }\rc^2
\eeq
for gravito-inertial waves ($s=1,2,\dots$),
\beq
\Lambda_{sm}\lp \nu \rp = \lp \nu -m  \rp^2
\eeq
for Yanai waves ($s=0$),
\beq
\Lambda_{sm}\lp \nu \rp = m^2 \frac{2m \nu}{2m \nu +1}
\eeq
for Kelvin waves ($s=-1$), and
\beq
\Lambda_{sm}\lp \nu \rp = \lc \frac{1}{2} \nu \lp 2s+1 \rp - \frac{1}{2}
\sqrt{\nu^2 \lp 2s+1 \rp ^2 -4 \lp m\nu - m^2 \rp }\rc^2
\eeq
for Rossby waves ($s=1,2,\dots$).
These asymptotic values are shown in Fig.~\ref{fig:lambda} and are given
for $\nu > 2$ for gravito-inertial waves, Yanai waves, and Kelvin waves and for
$\nu > 20$ for Rossby waves. As $\nu$ increases,
so does the agreement between the exact and asymptotic value of $\Lambda$.

For gravito-inertial, Yanai and Rossby waves,
eigenfunctions are given by
\beqan
\lefteqn{\Theta \lp \zeta \rp = \frac{\lp \sqrt{\Lambda} \nu \rp
^{1/2}}{\Lambda -m^2}} \\
&& \times \lc s \lp \frac{m}{\sqrt{\Lambda}}+1 \rp H_{s-1}\lp \zeta \rp +
\frac{1}{2} \lp \frac{m}{\sqrt{\Lambda}}-1 \rp H_{s+1} \lp \zeta \rp \rc e^{-\zeta^2/2}
\nonumber
\eeqan
\beq
\widehat{\Theta} = H_s \lp \zeta \rp e^{-\zeta^2/2}
\eeq
\beqan
\lefteqn{\widetilde{\Theta} \lp \zeta \rp = m\frac{\lp \sqrt{\Lambda} \nu \rp
^{1/2}}{\Lambda-m^2}} \\
&& \times \lc s \lp \frac{\sqrt{\Lambda}}{m}+1 \rp H_{s-1}\lp \zeta \rp +
\frac{1}{2} \lp \frac{\sqrt{\Lambda}}{m}-1 \rp H_{s+1} \lp \zeta \rp \rc e^{-\zeta^2/2}
\nonumber
\eeqan
with $\zeta \equiv \cos \theta \lp \sqrt{\Lambda} \nu \rp ^{1/2}$
and where $H_s$ is the Hermite polynomial of order $s$.
For the Kelvin waves, we get
\beq
\Theta \lp \eta \rp = e^{-\eta^2/2}
\eeq
\beq
\widehat{\Theta} \lp \eta \rp = - \frac{1}{\lp -m\nu\rp ^{1/2}}
\frac{m^2}{2m\nu+1} \eta e^{-\eta^2/2}
\eeq
\beq
\widetilde{\Theta} \lp \eta \rp = - m \lp \frac{\eta^2}{2m\nu+1} 
+1 \rp \eta e^{-\eta^2/2}
\eeq
with $\eta\equiv \cos \theta\lp -m\nu \rp ^{1/2}$
(Townsend~2003).

\subsection{Radial dependency \label{sec:radial}}

The formalism used to treat the wave radial dependency is not modified compared
to the case of the non-rotating star. In the anelastic approximation $\sigma^2 \ll N^2$,
the adiabatic radial displacement $\xi_r(r)$ obeys
\beq
\frac{\diff^2 \psi}{\diff r^2} + \lp \frac{N^2}{\sigma ^2} -1 \rp
\frac{\Lambda}{r^2} \psi =0
\label{radial}
\eeq
with $\psi\lp r \rp = \sqrt{\rho} r^2 \xi_r\lp r \rp$, 
$N^{2} = N_{T}^{2} + N_{\mu}^{2} = 
\frac{g\delta}{H_{p}} (\nabla_{ad}-\nabla) + \frac{g\varphi}{H_{p}}\nabla_{\mu}$
and where we neglected the righthand side term (\cf Press~1981).
The radial wavenumber is still
\beq
k_r^2 = \lp \frac{N^2}{\sigma ^2} -1 \rp \frac{\Lambda}{r^2}.
\label{eq:kr}
\eeq
However, in contrast to the non-rotating case, the horizontal wave 
number\footnote{In the absence of
rotation, Legendre polynomials follow
$$
r^2 \nabla _h^2 \Theta = -\Lambda \Theta,
$$
but this is no longer the case when rotation is added (see Eq.~\ref{laplace}).}
is no longer given by $k_h^2 = \Lambda/r^2$.
When horizontal gradients are required, 
an equivalent value has to be taken that we define here as
\beq
\lambda_{sm}^2 \lp \nu \rp = \frac{ \moylc{ r^2 \nabla_h^2
\Theta_{sm} \lp \nu \rp } } { \moylc{ \Theta_{sm} \lp \nu \rp } }
\label{equivalent}
\eeq
(see Fig.~\ref{fig:lambda} bottom), where $\moyl{\dots}$ denotes latitudinal averages.
In the absence of rotation, we recover $\lambda_{sm}=\Lambda_{sm}$. 
We may define an `equivalent' horizontal wave 
number that is now given by $k_h^2 = \lambda/r^2$, and we have $k_r^2 \gg k_h^2$ 
as in the non-rotating case (except when $\Lambda \to 0$).

We now apply the WKB method to solve Eq.~(\ref{radial}). For a wave
propagating towards the surface, we write 
$\psi \lp r \rp = C \lp r \rp e^{-i \int{dr' k_r}}$ and get
\beq
\xi_r (r) \propto \left| k_r \right| ^{-1/2} \rho^{-1/2} r^{-2} e^{-i \int{dr' k_r}}.
\eeq
Assuming that wave damping is dominated by heat diffusion and that this term remains
small, the radial displacement 
becomes $\xi_r (r) e^{-\tau(r)/2}$ with 
\beq
\tau(r) = \int_{r_{c}}^{r} dr' \: \frac{\Lambda^{3/2}}{r'^{3}} \: 
\frac{K N N_{T}^{2}}{\sigma^{4}} \sqrt{\frac{N^{2}}{N^{2}-\sigma^{2}}}
\label{eq:tau}
\eeq
(Zahn et al.~1997).

\section{Angular momentum transport \label{sec:angmom}}

\begin{figure}
\centerline{
\psfig{figure=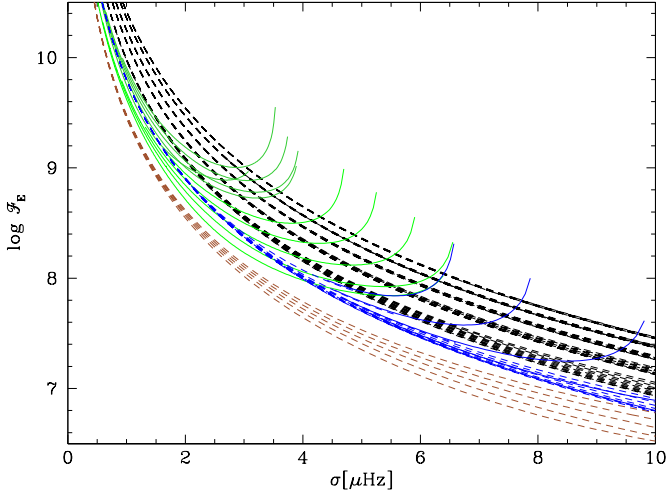,height=7cm,angle=-90}
}
\caption{Energy flux in IGWs at the base of radiative envelope in a
$3\,M_\odot$ main sequence star with $X_c=0.5$. {\bf Black:} Gravito-inertial waves {\bf Blue:} Yanai waves
{\bf Brown:} Kelvin waves {\bf Green:} Rossby waves. Continuous lines correspond
to numerical solutions and dashed lines,
to asymptotic solutions (see \S\,\ref{sec:asym}). 
\label{fig:fluxE}}
\end{figure}

\subsection{Excitation}

The treatment of wave excitation is more complex in the case of the rotating
star. Indeed, the traditional approximation, which requires 
$\sigma^2, (2\Omega)^2 \ll N^2$, is no longer valid in the convection zone; 
thus, formalisms that convolve the wave eigenfunctions with entropy
perturbations and Reynolds stresses in the convection zone (such as those by
Goldreich, Murray \& Kumar~1994, or Balmforth~1992) require major
transformations (Belkacem et al.~in prep.) For that reason, we resort to
adapting the Garc\'\i a L\'opez \& Spruit~(1991) formalism to our problem.

In that model, waves are excited at the 
boundary of the radiative zone by the
interface deformation caused by convective eddies. The eddies mean energy density
is transfered to waves of the same frequency and with $k_h \le k_{\rm eddy}$
according to
\beq
\frac{1}{2} \rho \moyh{ {\rm v}^2\lp\sigma \rp } =
\lp \frac{k_h}{k_{\rm eddy}} \rp ^2 \frac{1}{2} \rho \moyh{ {\rm v}_{\rm eddy}^2\lp
\sigma \rp }
\eeq
(Garc\'\i a L\'opez \& Spruit~1991), where $\moyh{\dots}$ denotes
horizontal averages. The factor $k_h/k_{\rm eddy}$ reflects
the reduced efficiency in the production of small $k_h$ waves caused by the
stochastic nature of turbulent eddies.
Assuming a turbulent Kolmogorov spectrum, we have $L_{\rm eddy}\propto 
{\rm v}_{\rm eddy}^3$ and $L_{\rm max} \le L_c$ with $L_c$ the mixing length.
The spectral distribution becomes
\beq
k_{\rm eddy}^2 \lp \sigma \rp = k_c^2 \lp \frac{\sigma}{\sigma_c} \rp ^3
\eeq
\beq
{\rm v}_{\rm eddy}^2 \lp \sigma \rp = {\rm v}_c^2 \lp \frac{\sigma}{\sigma_c} \rp ^{-1}
\eeq
with $\sigma \ge \sigma_c$, $k_c=2\pi/L_c$ and where 
${\rm v}_c$ and $\sigma_c$
are the characteristic convective velocities and frequencies, respectively.
We get
\beq
\frac{1}{2} \rho \moyh{ {\rm v}^2\lp\sigma \rp } =
\frac{1}{2} \rho {\rm v}_c^2 \lp \frac{k_h}{k_c} \rp ^2 
\lp \frac{\sigma}{\sigma_c} \rp ^{-4}.
\eeq
To proceed further, we assume that the horizontal dependence of IGWs at the
radiative boundary is given by Hough functions\footnote{This is formally
wrong since at this boundary we have $N^2\longrightarrow 0$, which
contradicts the traditional approximation. However, as $N^2$ rises
rapidly away from the boundary, and considering the inescapable presence of
a slight amount of overshooting, the approximation is acceptable.}.
Using the `equivalent' eigenvalue given by Eq.~(\ref{equivalent}), one then has
\beq
\frac{1}{2} \rho \moyh{ {\rm v}_{sm} ^2\lp\sigma \rp } =
\frac{\lambda_{sm}}{r_c^2 k_c^2} \, \frac{1}{2} \rho {\rm v}_c^2 
\lp \frac{\sigma}{\sigma_c} \rp ^{-4}.
\label{exc_mod}
\eeq
An alternative definition of the equivalent eigenvalue 
$\lambda = \moyl{ \left| r^2 \nabla_h^2 \Theta \right| } 
/ \moyl{ \left| \Theta \right| }$ leads to very similar results.
Following Mathis~(2005), one could also use a projection of the Hough functions onto 
spherical harmonics for which horizontal wavenumbers are well-defined.
However, in the case of fast rotation numerical problems are encountered with 
this last method because Legendre polynomials are not a good
basis for projecting Hermite polynomials. 

\begin{figure}
\centerline{
\psfig{figure=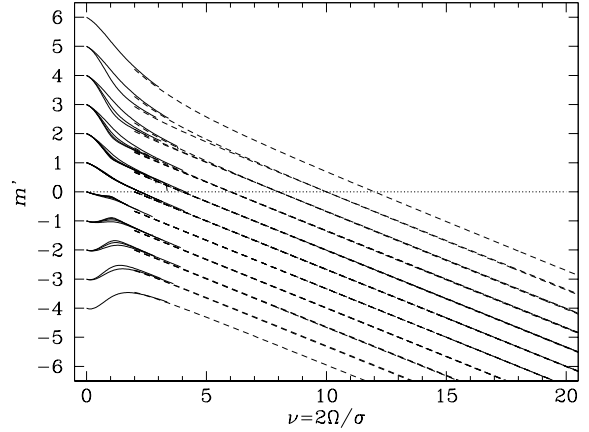,height=6cm,angle=-90}
}
\caption{Ratio $m'=-\frac{\sigma}{2} \frac{{\cal F}_J}{{\cal F}_E}$ for the 
gravito-inertial waves when the Coriolis acceleration is neglected
in the calculation of the angular momentum flux. 
Continuous lines correspond
to numerical solutions and dashed lines
to asymptotic solutions (see \S\,\ref{sec:asym}). 
\label{fig:m2faux}}
\end{figure}

\begin{figure*}
\centerline{
\psfig{figure=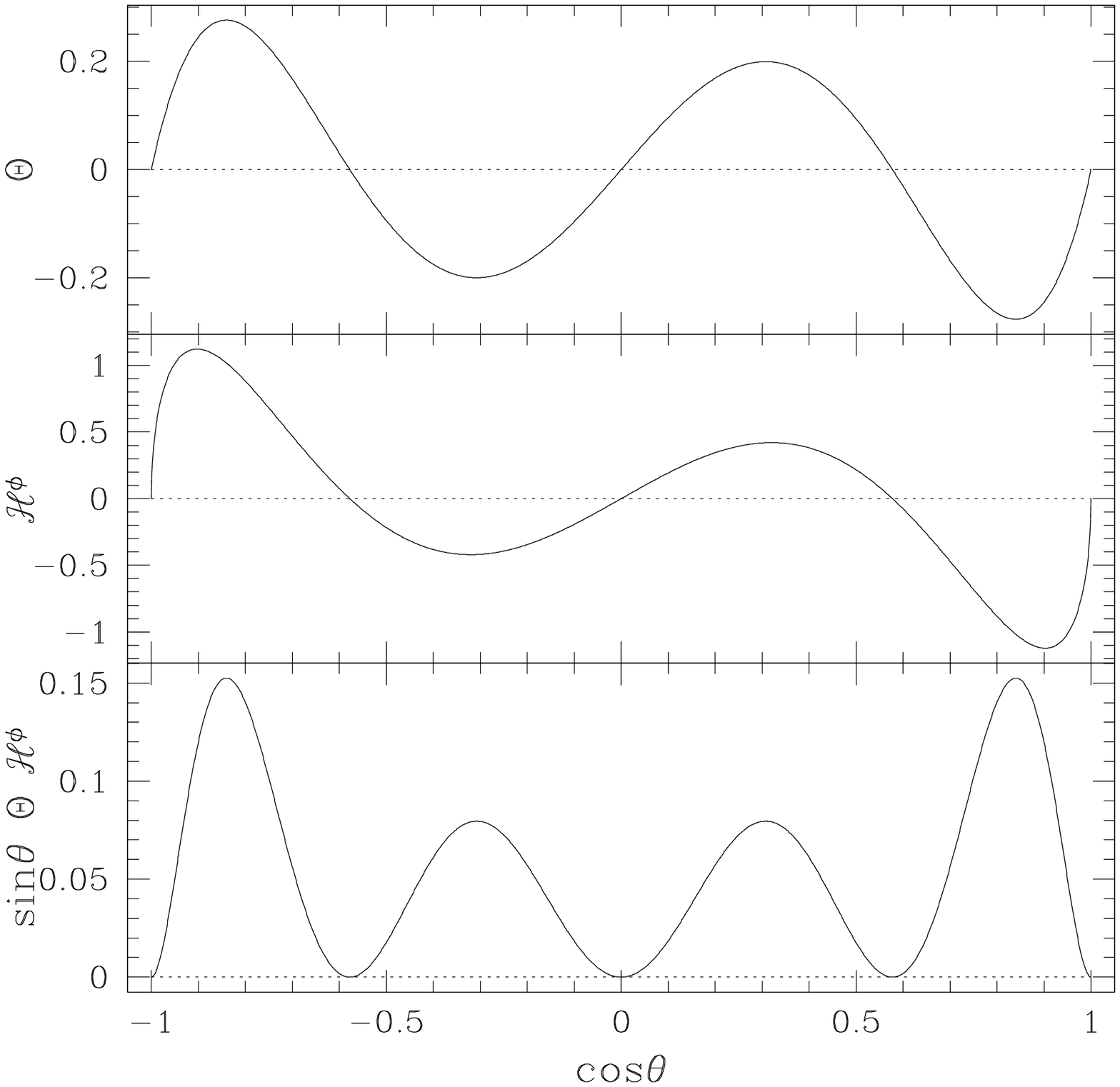,height=7.5cm}
\psfig{figure=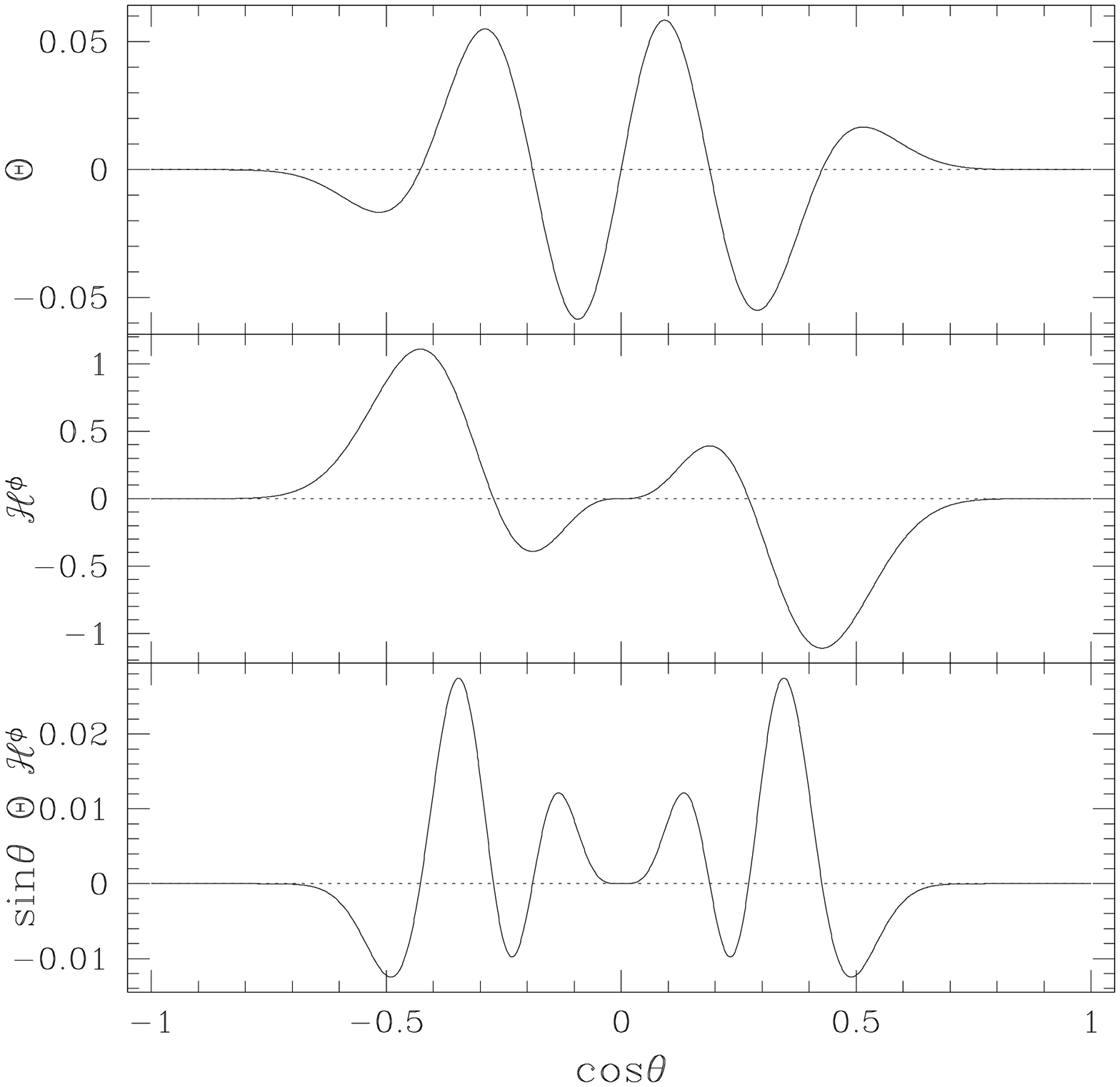,height=7.5cm}
}
\caption{{\bf Left:} 
Hough functions $\Theta_{sm}$ and
${\cal H}^\phi_{sm}$ for $s=4$, $m=2$ ($\ell = 5$), $\nu=0$,
and their product $\sin \theta\,\Theta {\cal H}^\phi$, which is
used for the angular momentum flux.
Normalization is given by $\diff \Theta /\diff \cos \theta=1$ at $\cos
\theta =0$. 
{\bf Right:} Equivalent Hough functions for $\nu \simeq 2$.
In the rotating case, the product $\sin \theta\,\Theta {\cal H}^\phi$
changes signs locally.
\label{fig:hough_flux}}
\end{figure*}

\subsection{Energy flux}

Angular momentum transport by IGWs is dominated by the low-frequency waves.
Here, we will thus adopt $\sigma^2 \ll N^2$ and use
$k_r^2 = \lp N/\sigma \rp ^2 \Lambda/r^2$.
The kinetic energy flux per unit frequency is given by
\beq
{\cal F}_E = \frac{1}{2} \rho \moyh{ {\rm v}^2 } {\rm v}_g,
\eeq
where the radial group velocity ${\rm v}_g$ is given by
\beq
{\rm v}_g = \frac{\diff \sigma}{\diff k_r} = -\frac{\sigma}{k_r}.
\label{vgroup}
\eeq
In the case of 
main sequence massive stars, we have $k_r<0$, corresponding to waves
traveling from the convective core 
to the surface. Using Eq.~(\ref{exc_mod}) 
to get the energy of a given mode, we obtain
\beq
{\cal F}_E \lp \sigma,s,m \rp = \frac{\rho \lambda_{sm}}{2\sqrt{\Lambda_{sm}}} 
\frac{{\rm v}_c^2 \sigma _c ^2}
{r_c k_c^2 N_c} \lp \frac{\sigma}{\sigma _c} \rp ^{-2},
\eeq
where $N_c$, which would formally be $0$ at the interface, is taken a fraction
of a pressure scale height into the radiative region.

These fluxes have been computed for a $3\,M_\odot$ main-sequence, 
population~I star calculated with the Geneva stellar evolution code 
and for a rotation velocity of $\Omega = 20\,\mu{\rm Hz}$
(Fig.~\ref{fig:fluxE}). This value, typical of massive stars,
corresponds to $\sim 40\%$ of the surface
critical velocity.

In the case of Rossby and retrograde Yanai waves, one can see that they appear with a significant
energy flux. This is caused by the divergence of the group velocity
when $\Lambda_{sm}$ is close to $0$ (see Eq.~\ref{vgroup}). This situation is
unphysical; however, it will be compensated for by a term in $\Lambda$ in the
calculation of the angular momentum flux (Eq.~\ref{eq:fluxJ}). Thus
we do not need to correct this in the present treatment. 

\begin{figure*}
\centerline{
\psfig{figure=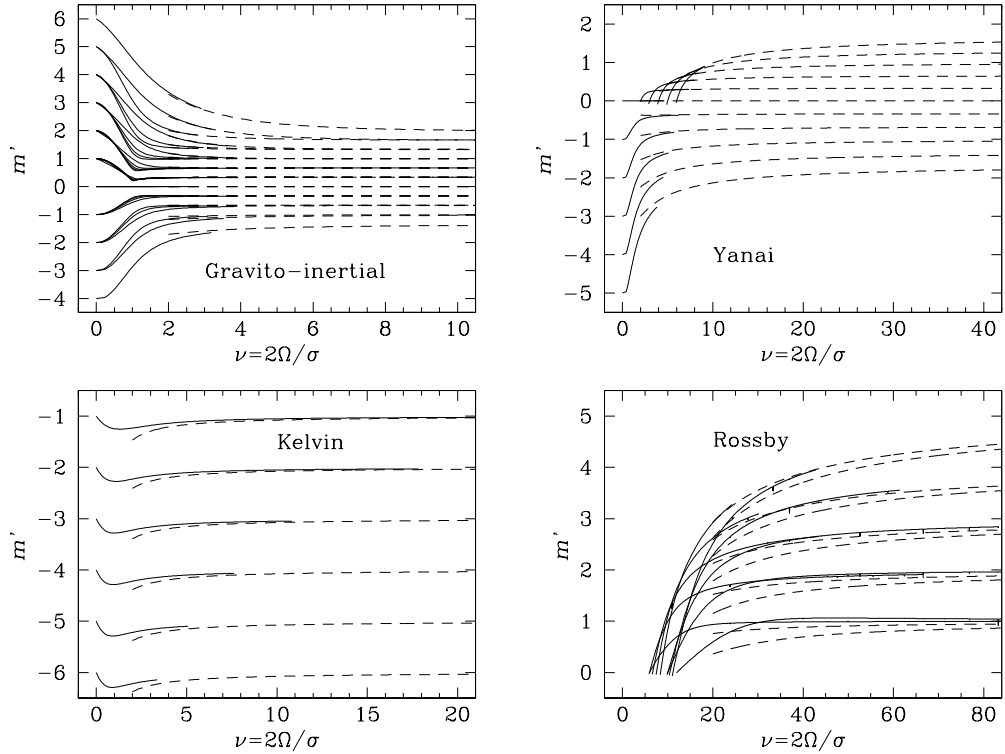,height=11cm,angle=-90}
}
\caption{Ratio $m'=-\frac{\sigma}{2} \frac{{\cal F}_J}{{\cal F}_E}$ for various modes.
{\bf Top left:} Gravito-inertial waves corresponding to $\ell \le 6$ 
{\bf Top right:} Yanai waves of order $m=-5,\dots, +3$.
{\bf Bottom left:} Kelvin waves of order $m\ge -6$.
{\bf Bottom right:} Rossby waves of order $s=1,2$, and $m\le 4$. 
Continuous lines correspond
to numerical solutions and dashed lines
to asymptotic solutions (see \S\,\ref{sec:asym}). 
\label{fig:m2}}
\end{figure*}

\subsection{Angular momentum flux}

We must now convert the energy flux into an angular momentum flux.
Following Zahn et al.~(1997), we write
\beq
{\cal F}_E = \frac{1}{2} \rho 
\moyh{ {\rm v}_r^2 + {\rm v}_\theta ^2 + {\rm v}_\phi ^2 } {\rm v}_g,
\eeq
with the real part of the velocity field $\vec{\rm v}=i\sigma \vec{\xi}$
\beqan
{\rm v}_r 
= A(r) \sin \lp \hspace*{-0.02cm} \sigma t - \hspace*{-0.09cm} 
\int _{r_c} ^ r \hspace*{-0.11cm} \diff r'
k_r \hspace*{-0.05cm}+ \hspace*{-0.05cm}m \phi \hspace*{-0.02cm}
\rp \Theta \lp \nu, \cos \theta \rp e^{-\tau(r)/2}
\nonumber
\eeqan
\beqan
{\rm v}_\theta 
= -\frac{rk_r}{\Lambda} A(r) \cos \lp \hspace*{-0.02cm} \sigma t - \hspace*{-0.09cm} 
\int _{r_c} ^ r \hspace*{-0.11cm} \diff r'
k_r \hspace*{-0.05cm}+ \hspace*{-0.05cm}m \phi \hspace*{-0.02cm}
\rp {\cal H}^\theta \lp \nu, \cos \theta \rp e^{-\tau(r)/2}
\nonumber
\eeqan
\beqan
{\rm v}_\phi 
= \frac{rk_r}{\Lambda}
A(r) \sin \lp \hspace*{-0.02cm} \sigma t - \hspace*{-0.09cm} 
\int _{r_c} ^ r \hspace*{-0.11cm} \diff r'
k_r \hspace*{-0.05cm}+ \hspace*{-0.05cm}m \phi \hspace*{-0.02cm}
\rp {\cal H}^\phi \lp \nu, \cos \theta \rp e^{-\tau(r)/2},
\nonumber
\eeqan
given by the WKB method for ${\rm v}_r$, and with help of the continuity equation 
in the Boussinesq approximation for ${\rm v}_\theta$ and ${\rm v}_\phi$.
The energy flux for a given mode becomes
\beqan
{\cal F}_E = -\frac{A^2(r)}{4}\rho \lp \moylc{ \Theta } +
\frac{r^2 k_r^2}{\Lambda^2} \lc 
\moylc{ {\cal H}^\theta } + \moylc{ {\cal H}^\phi }
\rc \rp \frac{\sigma}{k_r} e^{-\tau(r)}.
\nonumber
\eeqan
This equation is equivalent to the one obtained by Mathis~(2005).

In the absence of rotation, 
the angular momentum flux is expressed as (Zahn et al.~1997)
\beq
{\cal F}_J= \moyh{ \rho r \sin \theta \,{\rm v}_r {\rm v}_\phi },
\eeq
which becomes
\beq
{\cal F}_J= \frac{A^2(r)}{2} \rho r
\moyl{ \sin \theta \, \Theta \, {\cal H}^\phi } 
\frac{rk_r}{\Lambda} e^{-\tau(r)}
\eeq
for a given mode. 
When this simple formulation is applied to gravito-inertial waves, one finds
that the sign of angular momentum transport varies with $\nu$ (see Fig.~\ref{fig:m2faux}).
As the rotation parameter $\nu$ increases, the value $m'=-\frac{\sigma}{2} 
\frac{{\cal F}_J}{{\cal F}_E}$ diminishes
linearly and changes sign for $\nu=2m$.
This would imply that a {\em retrograde} wave ($m>0$) would, for fast enough rotation,
carry a {\em positive} angular momentum flux ($m'<0$).
This can be understood with the aid of Fig.~\ref{fig:hough_flux}. In the
absence of rotation, $\Theta = P_\ell^m$ and
${\cal H}^\phi=m P_\ell^m /\sin \theta$; their product always has
the sign of $m$. This is no longer the case when the Coriolis acceleration
is included. The integral $\moyl{ \sin \theta \, \Theta \, {\cal H}^\phi }$
may thus change sign. 

The solution to this paradox lies in the fact that, in a rotating
system, the angular momentum flux is actually given by
\beq
{\cal F}_J= \moyh{ \rho r \sin \theta \,{\rm v}_r \lp {\rm v}_\phi
+ 2 \Omega \cos \theta \xi_\theta \rp }
\eeq
(Jones~1967; Bretherton~1969). The first term in this equation corresponds to the
angular momentum flux across an Eulerian surface, and the second term to a flux
associated with a Lagrangian contribution to angular momentum (Bretherton~1969).
In that case, the actual angular momentum flux thus becomes
\beqan
{\cal F}_J= \frac{A^2(r)}{2} \rho r 
\lp \moyl{ \sin \theta \, \Theta \, {\cal H}^\phi }
- \nu \moyl{ \sin \theta \, \cos \theta \, \Theta \, {\cal H}^\theta } \rp
\frac{rk_r}{\Lambda} e^{-\tau(r)} \nonumber
\eeqan
for a given mode. The ratio between the kinetic energy and the angular
momentum flux is thus
\beq
\frac{{\cal F}_J}{{\cal F}_E} = -\frac{2 \Lambda}{\sigma} 
\frac{\moyl{ \sin \theta \, \Theta \, {\cal H}^\phi } 
- \nu \moyl{ \sin \theta \, \cos \theta \, \Theta \, {\cal H}^\theta } }
{ \moylc{ {\cal H}^\theta } + \moylc{ {\cal H}^\phi } },
\label{eq:fluxJ}
\eeq
where we use the approximation $r^2 k_r^2 \gg \Lambda ^2$. In the absence
of rotation, we get
\beq
\frac{{\cal F}_J}{{\cal F}_E} = -\frac{2 m}{\sigma}.
\eeq
We define an equivalent azimuthal number $m'$ such that 
\beq
m'\equiv -\frac{\sigma}{2} \frac{{\cal F}_J}{{\cal F}_E}
= \Lambda  \frac{\moyl{ \sin \theta \, \Theta \, {\cal H}^\phi } 
- \nu \moyl{ \sin \theta \, \cos \theta \, \Theta \, {\cal H}^\theta } }
{ \moylc{ {\cal H}^\theta } + \moylc{ {\cal H}^\phi } },
\eeq
which has been evaluated numerically and is shown in Fig.~\ref{fig:m2}.

In the case of gravito-inertial waves, as $\nu$ increases,
the value $m'$ rapidly converges
towards $m/3$, for all orders $s$. This implies that the symmetry that exists
between prograde and retrogrades waves is conserved in the rotating case\footnote{Note
however that excitation could be asymmetric.}. Yanai waves behave similarly, but
with a slower convergence rate.

In the case of the Kelvin waves, $m'$ varies only slightly with rotation and
remains close to $m$. Their angular momentum flux is always positive. 
Rossby waves appear with $m'\simeq 0$, and its value rises
slowly and tends towards $m$ with the increase in the rotation rate. Their angular
momentum flux is always negative.

\begin{figure}
\centerline{
\psfig{figure=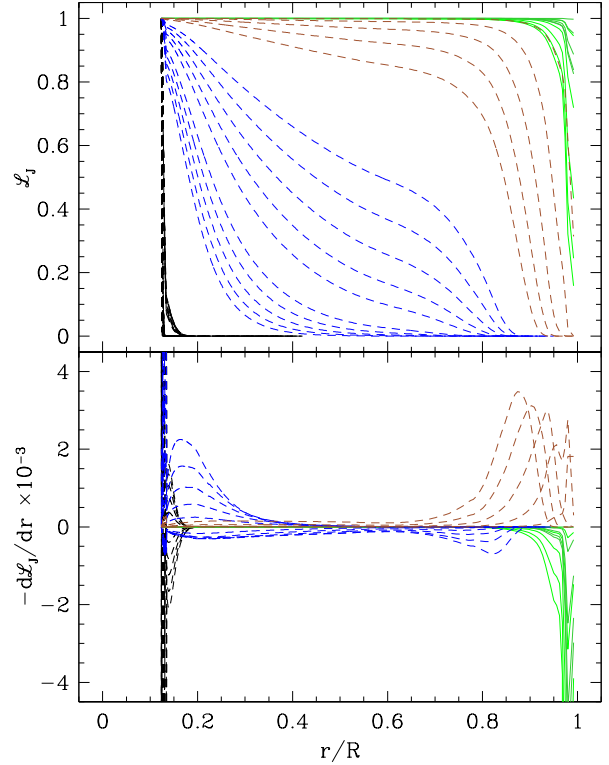,height=11cm}
}
\caption{Local angular momentum luminosity normalized to 1 at the base of
the radiative zone (top) and local deposition of angular momentum (bottom)
for waves with $\sigma=3\,\mu {\rm Hz}$ in a $3\,M_\odot$ 
main sequence star ($X_c=0.5$).
{\bf Black:} Gravito-inertial waves {\bf Blue:} Yanai waves
{\bf Brown:} Kelvin waves {\bf Green:} Rossby waves. Continuous lines correspond
to numerical solutions and dashed lines
to asymptotic solutions (see \S\,\ref{sec:asym}). 
\label{fig:lumloc}}
\end{figure}

\subsection{Angular momentum deposition}

The angular momentum distribution within the star evolves under the effect of the
damping of IGWs. When traveling inwards, each wave deposits its angular
momentum at the location where it is damped. In stars, the major source
of damping is thermal diffusivity (\cf \S\,\ref{sec:radial}), which is different
for each wave. We define the angular momentum luminosity
${\cal L}_J = 4\pi r^2 {\cal F}_J$,
which in the WKB approximation can be expressed as
\beq
{\cal L}_J \lp r \rp = 4\pi r^2 {\cal F}_J \lp r_c \rp e^{-\tau \lp r \rp}.
\eeq
Then, the local angular momentum evolves according to
\beq
\frac{\diff }{\diff t} \lp \frac{8\pi}{3} \rho \Omega r^4 \rp
= - \frac {\diff }{\diff r} \lp \int \diff \sigma \sum_{s,m} {\cal L}_J\lp \sigma,s,m,r \rp \rp
\eeq
(Zahn et al.~1997).

In this exploratory step, we examine {\em where} damping occurs for each
type of wave. Let us first look at Fig.~\ref{fig:lumloc}, which gives 
the local angular momentum luminosity and the angular momentum deposition 
$-\diff {\cal L}_J/\diff r$ of waves with a frequency 
$\sigma=3\,\mu {\rm Hz}$, for the $3\,M_\odot$ model of Fig.~\ref{fig:fluxE} 
(the rotation parameter is $\nu=13.3$). Since all waves shown
here have the same frequency, the location of the deposition is determined
by the eigenvalue $\Lambda$ (see Eq.~\ref{eq:tau}). The rapid fall in the local 
amplitude at the convective boundary (over $\delta r \simeq 1\%$) is caused 
by the mean molecular weight gradient that has been left behind the regressing core,
and damping increases in the outer region with the thermal diffusivity $K$.

\section{Discussion}

In this paper, we have examined the transport of angular momentum by low-frequency
waves excited by core convection. The traditional approximation was used
to evaluate the impact of the Coriolis acceleration on waves. We find two main
effects: \\
$-$ the horizontal structure, and hence the amount of angular momentum carried
by a wave, is modified; \\
$-$ new types of waves appear.

In the case of gravito-inertial waves, the main effect of the Coriolis acceleration
is to confine the wave towards the equator. This reduces the total angular momentum
carried by a wave of a given amplitude and frequency (Fig.~\ref{fig:m2}). 
As rotation increases, a saturation exists in this reduction, which
tends towards $m'=-\frac{\sigma}{2} \frac{{\cal F}_J}{{\cal F}_E}=\frac{m}{3}$.
Rotation also increases their radial wave number (see Eq.~\ref{eq:kr} and 
Fig.~\ref{fig:lambda}), and hence their damping. These waves are thus deposited 
very close to the convection 
core (see Fig.~\ref{fig:lumloc}). If the asymmetry
in the excitation of prograde and retrograde waves is not too strong, we expect
that the damping of these waves could produce a shear layer oscillation (SLO) similar
to the one present in slowly rotating stars (Talon \& Charbonnel~2005 and
references therein).

The Yanai waves, whose main restoring force is also gravity for high values
of the rotation parameter $\nu = 2\Omega/\sigma$, have the same limit value for
$m'$. Their eigenvalues $\Lambda$ are, however, lower so they are damped
farther from the convection 
core (Fig.~\ref{fig:lumloc}) and over a larger portion
of the star. These most certainly do not produce a second SLO but could generate a
local shear in the interior (corresponding to the stationary solution of
Kim \& MacGregor~2001).

The other two types of waves, namely Kelvin and Rossby waves, show a somewhat different 
behavior. For these waves, the main restoring force is the conservation of
vorticity, combined with stratification in the first case and in the second, 
combined with curvature. These two types of waves have the same limiting value 
$m'=m$, and their eigenvalues $\Lambda$ are lower than those of both gravito-inertial
and Yanai waves. As a result, they are damped much closer to the stellar surface 
(Fig.~\ref{fig:lumloc}). These waves could create a strong shear in that region and 
could thus induce a large amount of mixing close to the stellar surface.
Complete dynamical simulations remain to be completed to verify these conjectures.


\end{document}